\documentclass{amsart}

\usepackage{amsmath}
\usepackage{amsmath,amsfonts}
\usepackage[english]{babel}
\usepackage{graphicx}

\newcommand{\kr}{Z}
\newcommand{\Tplus}{T_{+}}
\newcommand{\Tminus}{T_{-}}
\newcommand{\argmax}{\operatorname*{arg\ max}}

\newcommand{\lacto}{\emph{Lactobacillus}}
\newcommand{\liners}{\emph{L.~iners}}
\newcommand{\lcrisp}{\emph{L.~crispatus}}
\newcommand{\sneathia}{\emph{Sneathia}}
\newcommand{\prevotella}{\emph{Prevotella}}

\newcommand{\pplacer}{\textsf{pplacer}}
\newcommand{\guppy}{\textsf{guppy}}
\newcommand{\pplacerurl}{\texttt{http://matsen.fhcrc.org/pplacer/}}

\newcommand{\forarxiv}[1]{#1}
\newcommand{\notforarxiv}[1]{}

\newcommand{\FIGClusterMunkresRef}{S1}
\newcommand{\FIGEdgePCAPlotPhRef}{S2}

\newcommand{\FIGPCOne}{
\begin{figure}[ht]
\begin{center}
  \forarxiv{\includegraphics[height=10cm]{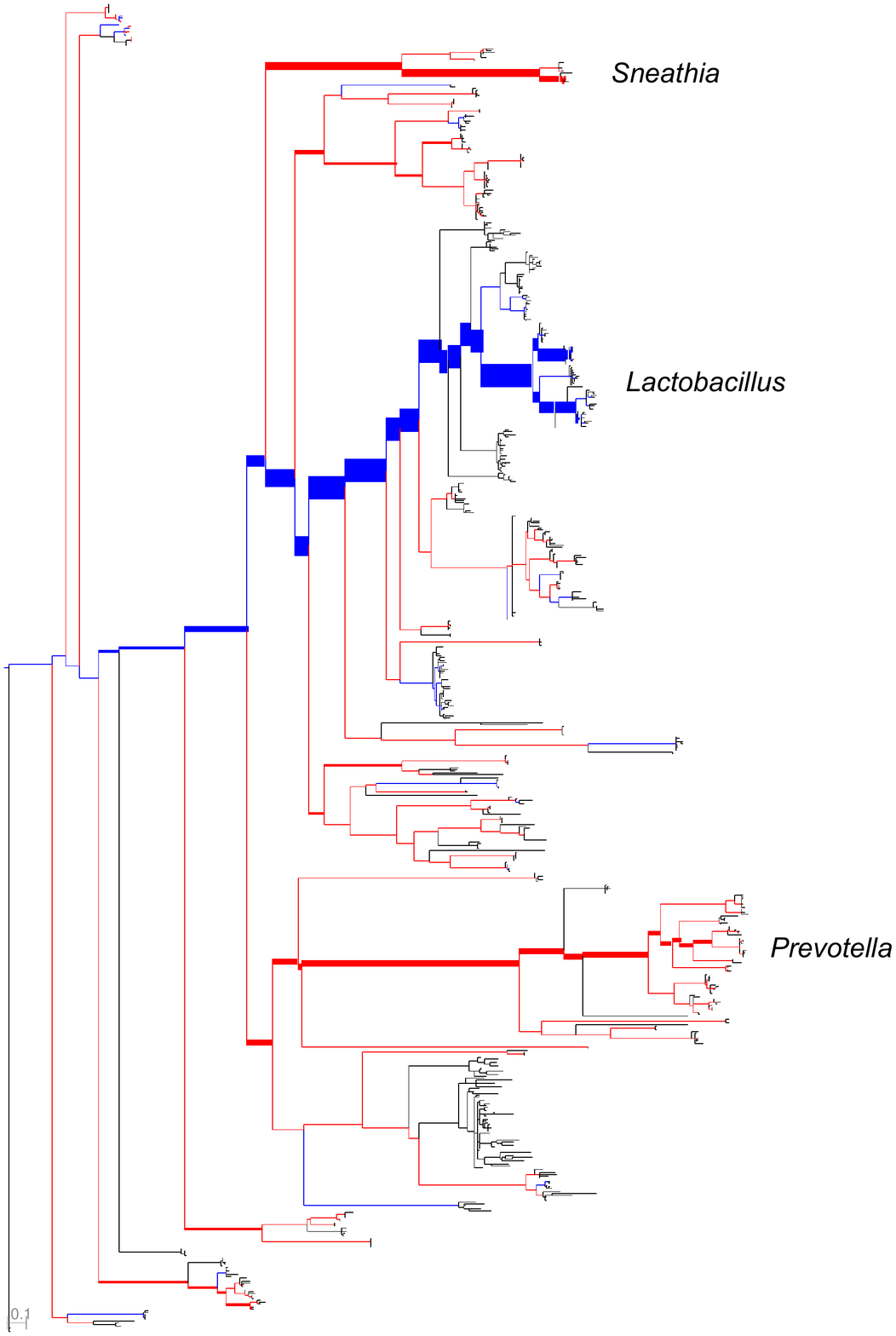}}
\end{center}
\caption{
The first principal component for the vaginal data set, representing about 66 percent of the variance.
The edges across which maximal between-sample heterogeneity is found are those leading to the \lacto\ clade and those leading to the Sneathia and Prevotella clade.
This axis appears to represent the bacterial vaginosis axis, as \sneathia\ and \prevotella\ are associated with bacterial vaginosis, while \lacto\ is associated with its absence.
}
\label{FIGPCOne}
\end{figure}
}

\newcommand{\FIGPCTwo}{
\begin{figure}[ht]
\begin{center}
  \forarxiv{\includegraphics[height=7cm]{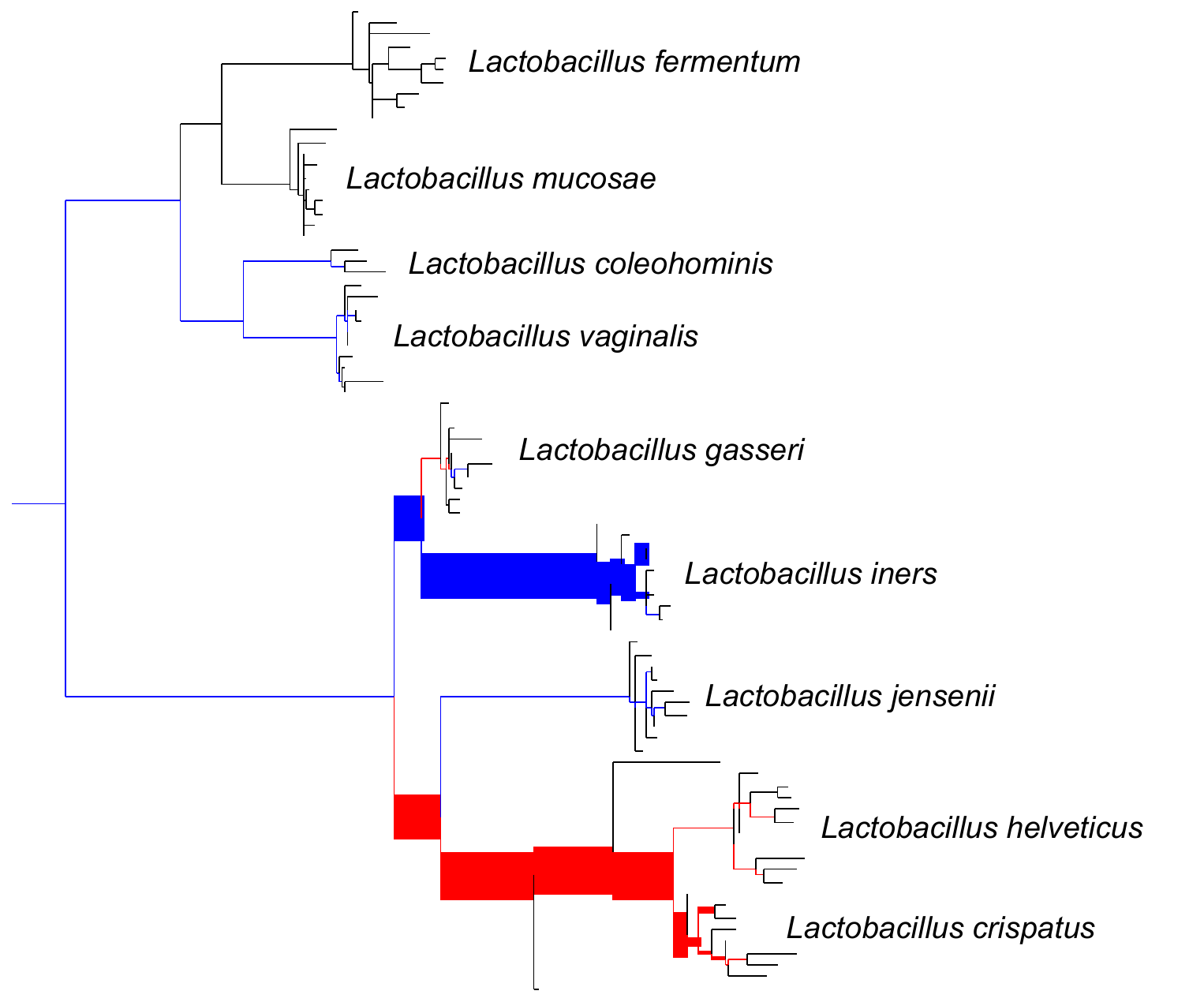}}
\end{center}
\caption{
The second principal component for the vaginal data set, representing about
19 percent of the variance (low-weight regions of the tree are excluded from the figure).
The edges across which maximal between-sample heterogeneity is found are those between two different \lacto\ clades: \liners\ and \lcrisp.
Thus, the second important ``axis'' appears to correspond to the relative levels of these two species.
}
\label{FIGPCTwo}
\end{figure}
}

\newcommand{\FIGEdgePCAPlot}{
\begin{figure}[ht]
\begin{center}
  \forarxiv{\includegraphics[width=9cm]{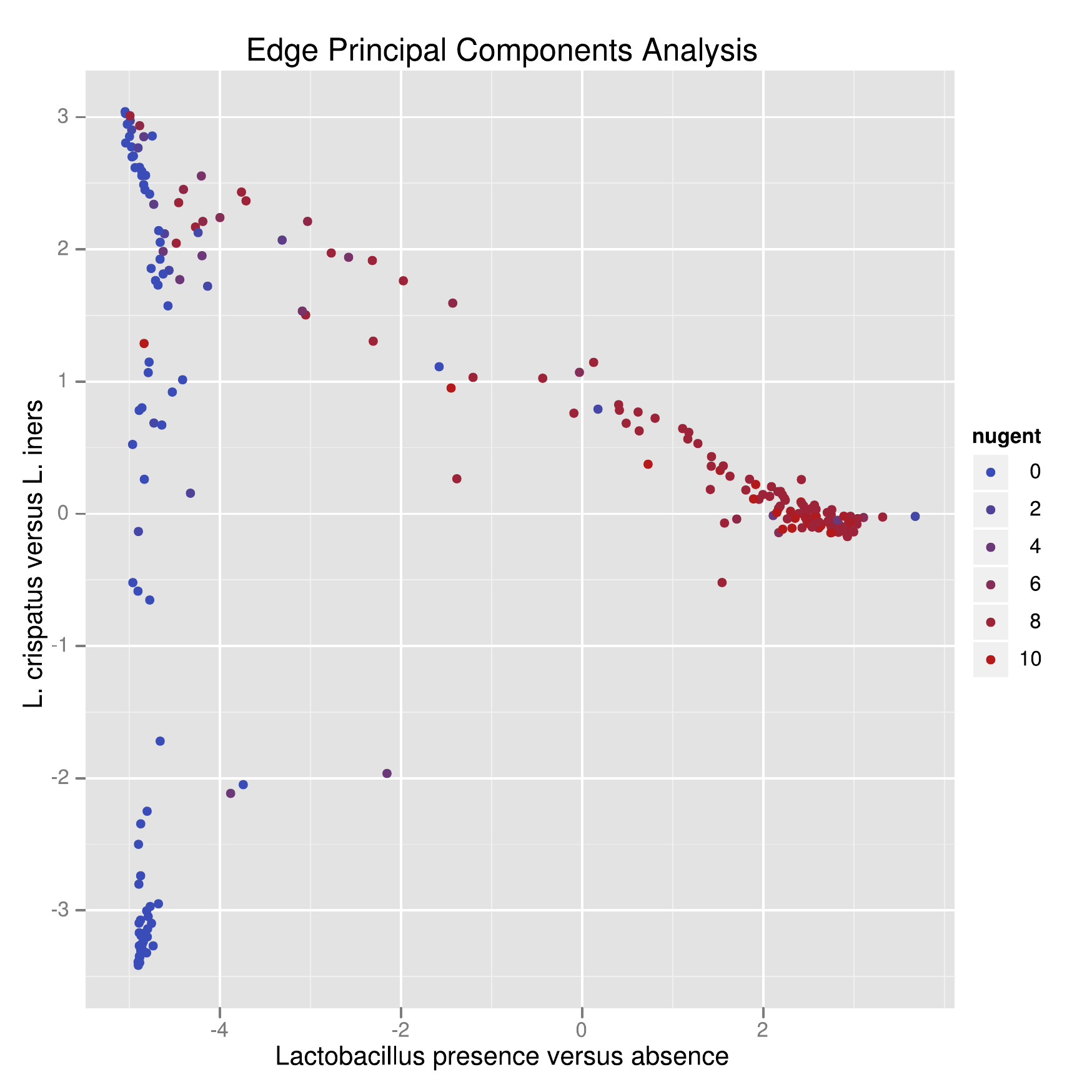}}
\end{center}
\caption{
Edge principal components analysis (edge PCA) applied to the vaginal data set.
The axes for the edge principal components plot are described in Figures~\ref{FIGPCOne} ($x$-axis) and \ref{FIGPCTwo} ($y$-axis).
The Nugent score is a diagnostic score for bacterial vaginosis, with high score indicating bacterial vaginosis.
}
\label{FIGEdgePCAPlot}
\end{figure}
}

\newcommand{\FIGClassicalPCAPlot}{
\begin{figure}[ht]
\begin{center}
  \forarxiv{\includegraphics[width=9cm]{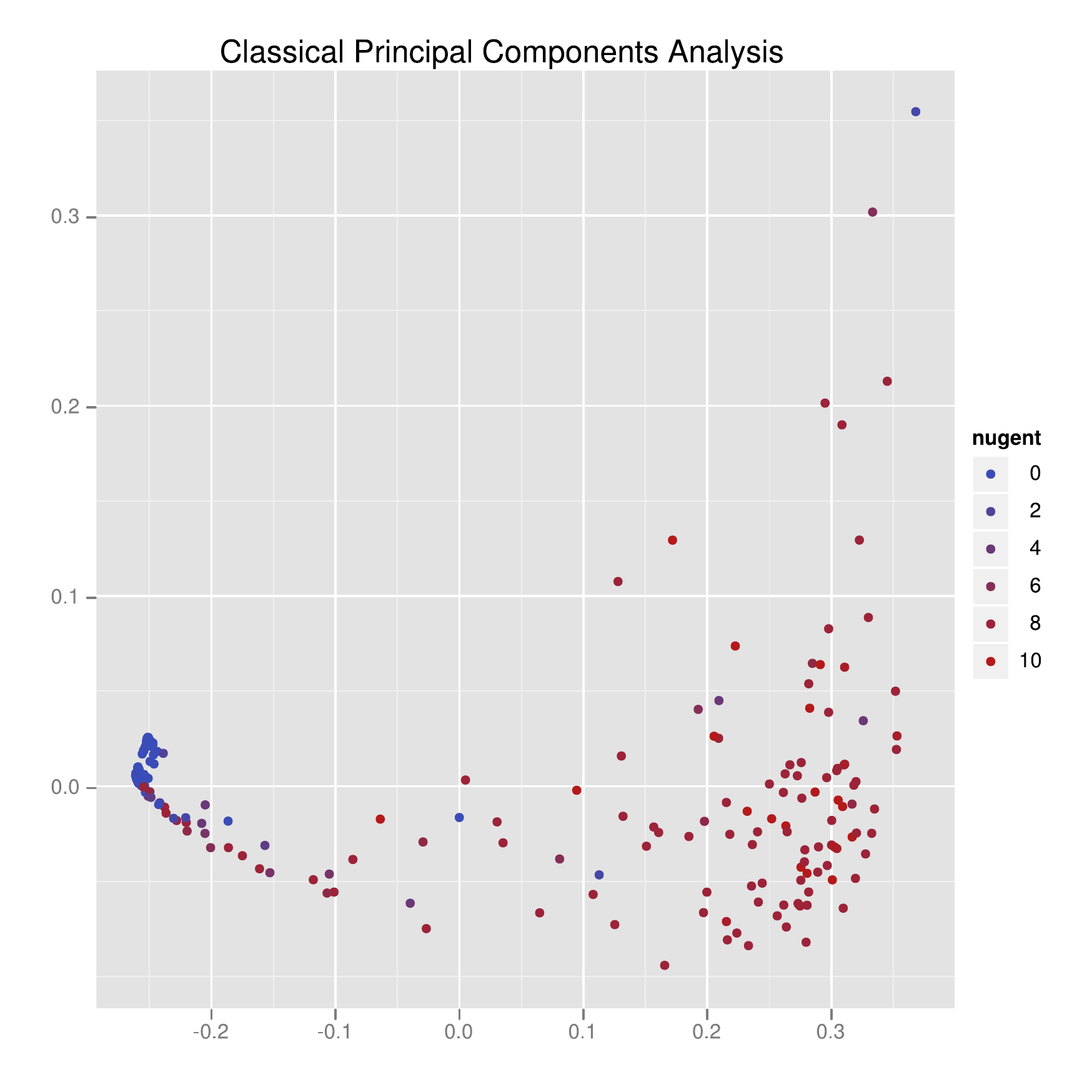}}
\end{center}
\caption{
Classical principal components analysis applied to the vaginal data set.
}
\label{FIGClassicalPCAPlot}
\end{figure}
}

\newcommand{\FIGClusterTrees}{
\begin{figure}[ht]
\begin{center}
  \forarxiv{\includegraphics[width=10cm]{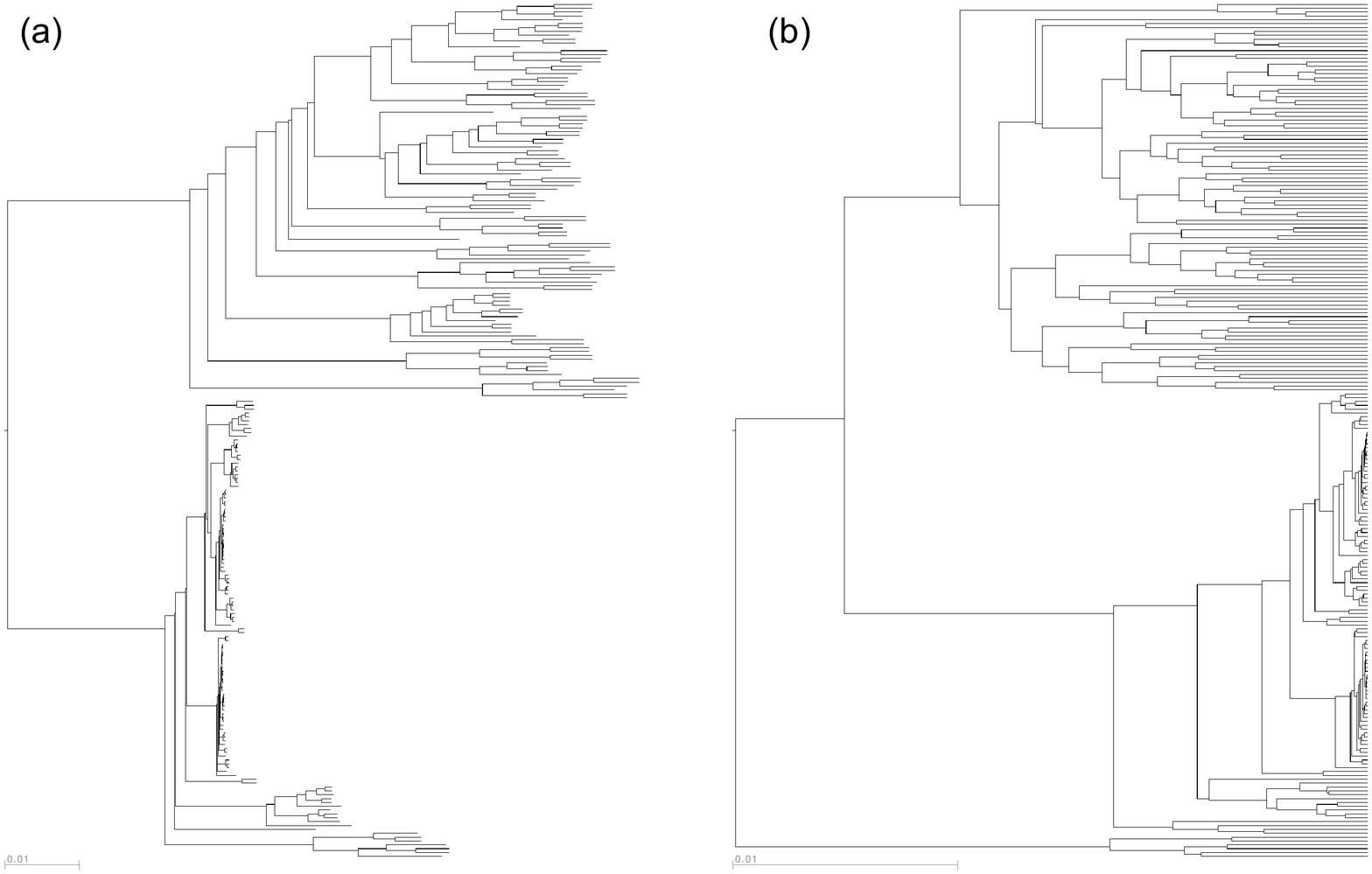}}
\end{center}
\caption{
The results of (a) squash clustering and (b) UPGMA as applied to the vaginal data set.
The labels are not shown and they do not appear in the same order on the two trees. For a comparison of labeled trees, see Supplementary Figure~\FIGClusterMunkresRef.
}
\label{FIGClusterTrees}
\end{figure}
}

\newcommand{\FIGSquash}{
\begin{figure}[ht]
\begin{center}
  \forarxiv{\includegraphics[width=10cm]{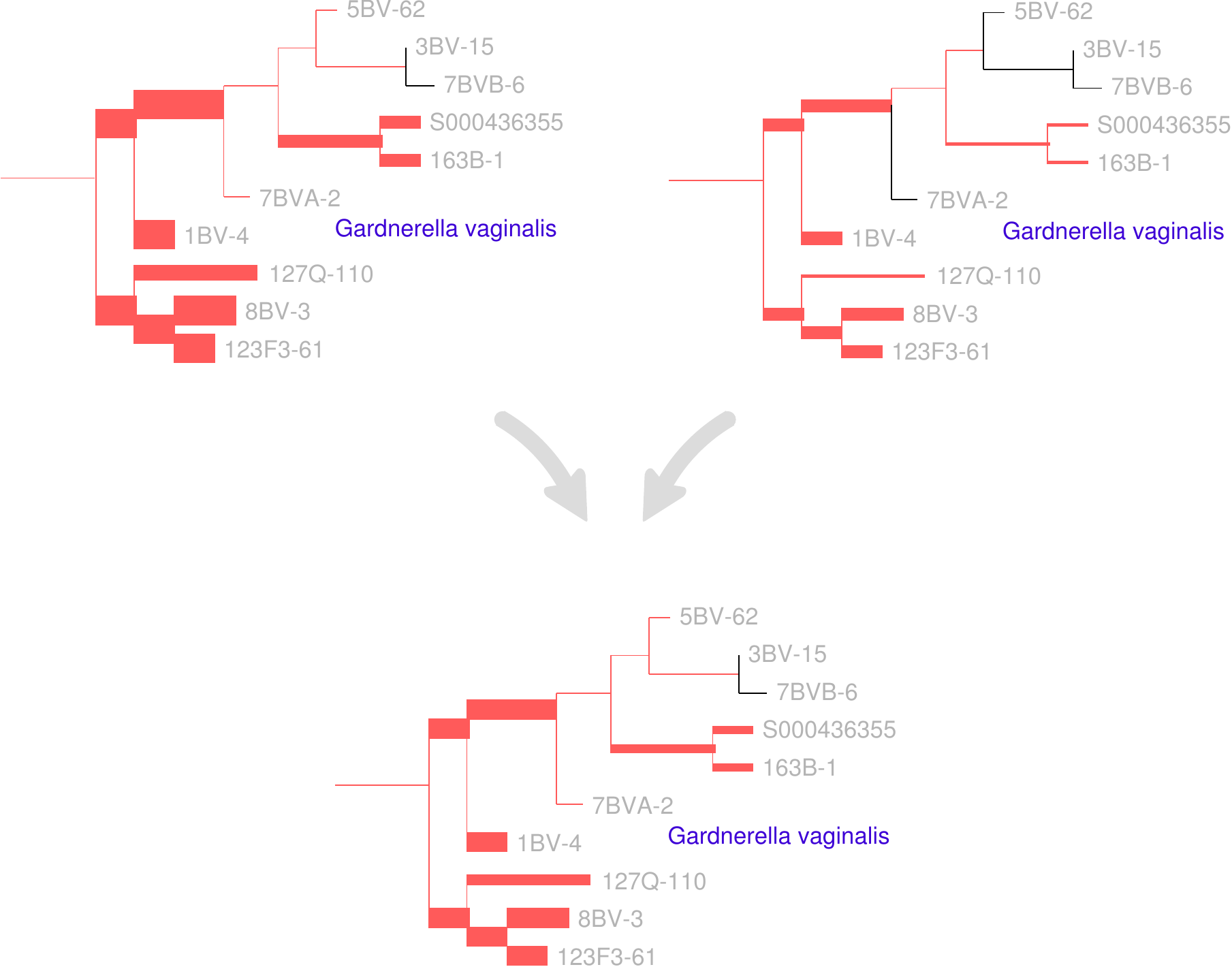}}
\end{center}
\caption{
A visual depiction of the squash clustering algorithm.
When two clusters are merged in the squash clustering algorithm, their mass distributions are combined according to a weighted average as described in the text.
The edges of the reference tree in this figure are thickened in proportion to the mass distribution (for simplicity, just a subtree of the reference tree is shown here).
In this example, the lower mass distribution is an equal-proportion average of the upper two mass distributions.
Similarities between mass distributions, such as the similarity seen between the two clusters for the \emph{G. vaginalis} clade shown here, are what cause clusters to be merged.
Such similarities between internal nodes can be visualized for the squash clustering algorithm; the software implementation produces such a visualization for every internal node of the clustering tree.
}
\label{FIGSquash}
\end{figure}
}

\newcommand{\FIGsquashsim}{
\begin{figure}[ht]
\begin{center}
  \forarxiv{\includegraphics[width=5.3in]{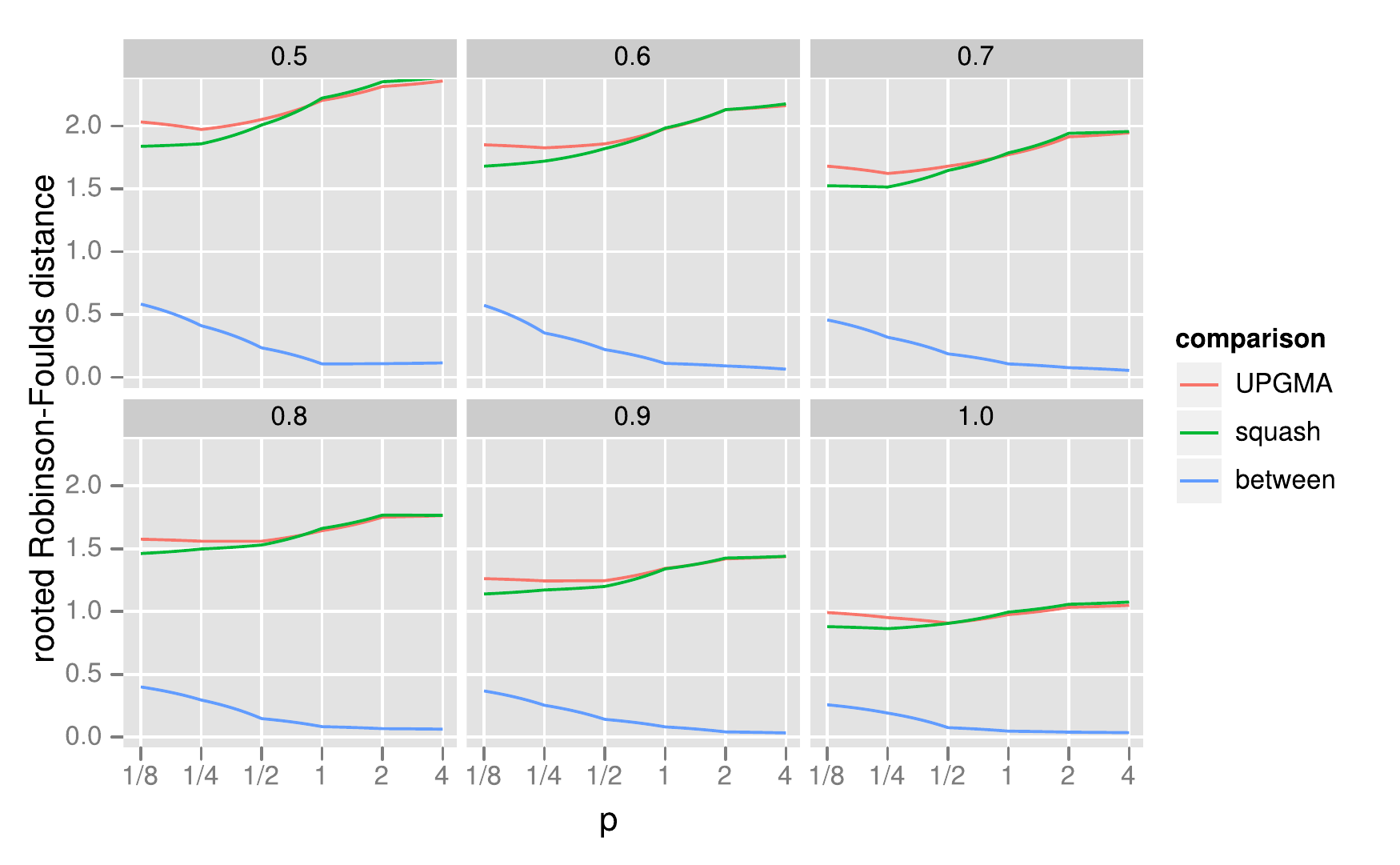}}
\end{center}
\caption{
The results of the cluster accuracy simulation experiment using the rooted
Robinson-Foulds (RF) metric.
This graphic shows very similar levels of topological accuracy for squash
clustering and UPGMA, as well as high similarity between the topology returned
by the two methods.
The figure is divided into panels by the level of reconstructability parameter
$r_t$ as described in the text (a larger $r_t$ implies easier reconstruction).
The $x$-axis is the value of $p$ for the $Z_p$ distance as described in \eqref{EQKR}.
The $y$-axis is the rooted Robinson-Foulds distance: for the ``squash'' and ``UPGMA'' lines
it is the distance between the reconstructed tree and the original tree using
these two algorithms (lower is more accurate), while the ``between'' line shows
the distance between the result for the two clustering algorithms (lower is
more similar).
Note that the maximum rooted RF distance between two trees with six taxa is four.
}
\label{FIGsquashsim}
\end{figure}
}

\newcommand{\FIGClusterMunkres}{
\begin{figure}[ht]
\begin{center}
  \forarxiv{\includegraphics[height=4.5in]{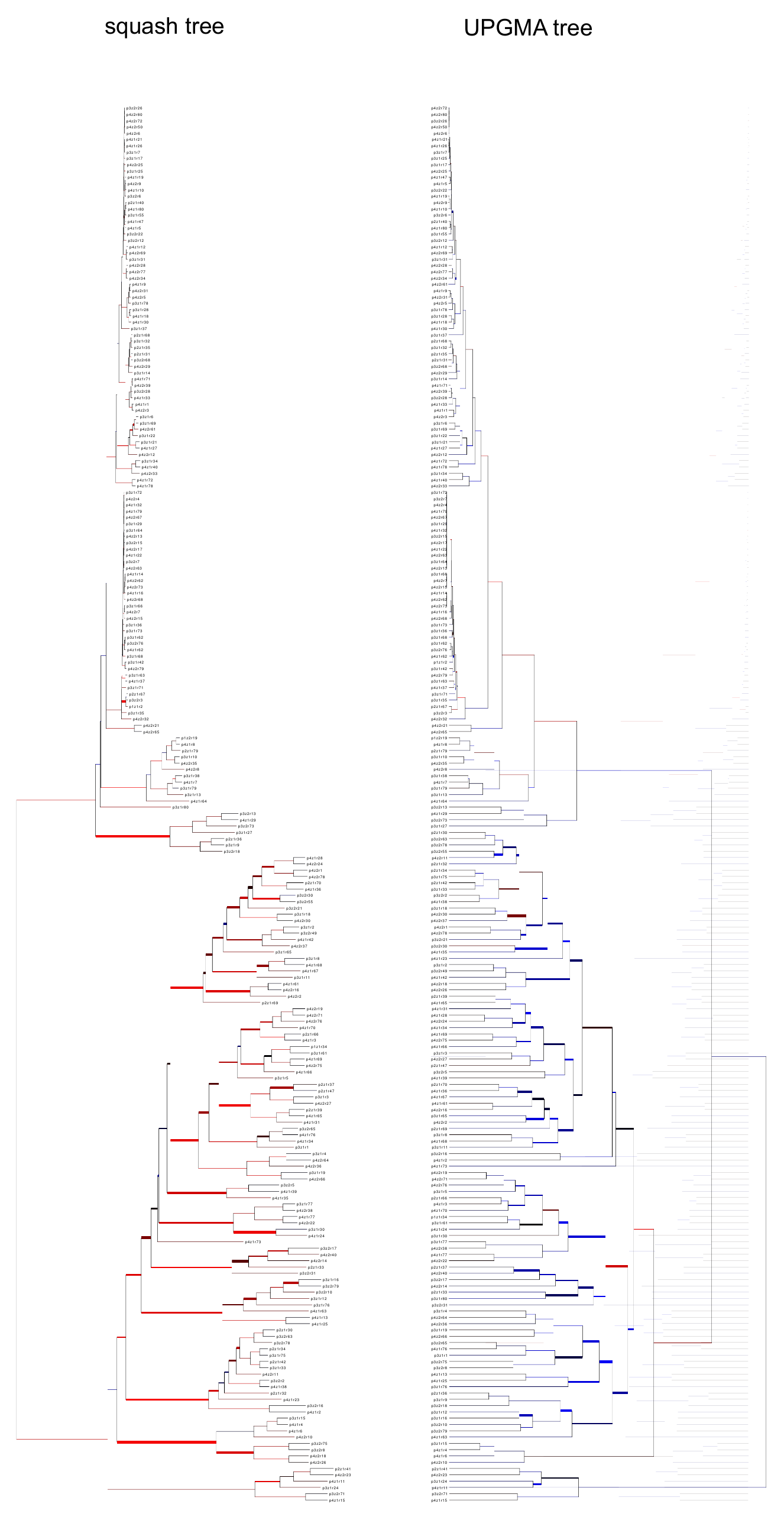}}
\end{center}
\caption{
A comparison of the clustering results for the vaginal data set using the software of \cite{NyeEaComparingTwoTrees06}.
The software uses the Hungarian (a.k.a. Munkres) algorithm to find an optimal one-to-one matching between edges of the trees minimizing differences in a topological score between pairs of matched branches as follows.
Given two trees $T$ and $S$ on the same samples, let $\Sigma (T)$ and $\Sigma (S)$ be the bipartitions of the samples induced by cutting the edges of $T$ and $S$.
For two bipartitions $i$ and $j$, one associates an ``agreement score'' $s(i,j)$ describing the proportion of shared elements between the sides of the bipartitions.
The algorithm finds a one-to-one matching between $\Sigma (T)$ and $\Sigma (S)$ that minimizes the total agreement score between matched bipartitions.
Each tree is drawn in a way which shows the agreement scores: a thick branch represents an edge which has a low agreement score with its partner in the matching.
The program arranges the trees such that matched edges are close to one another on the tree.
Branches shown in red mean the colored branch is longer than the branch in the other tree, while those in blue are opposite; the intensity of the color indicates the degree of this difference.
}
\label{FIGClusterMunkres}
\end{figure}
}

\newcommand{\FIGEdgePCAPlotPh}{
\begin{figure}[ht]
\begin{center}
  \forarxiv{\includegraphics[width=9cm]{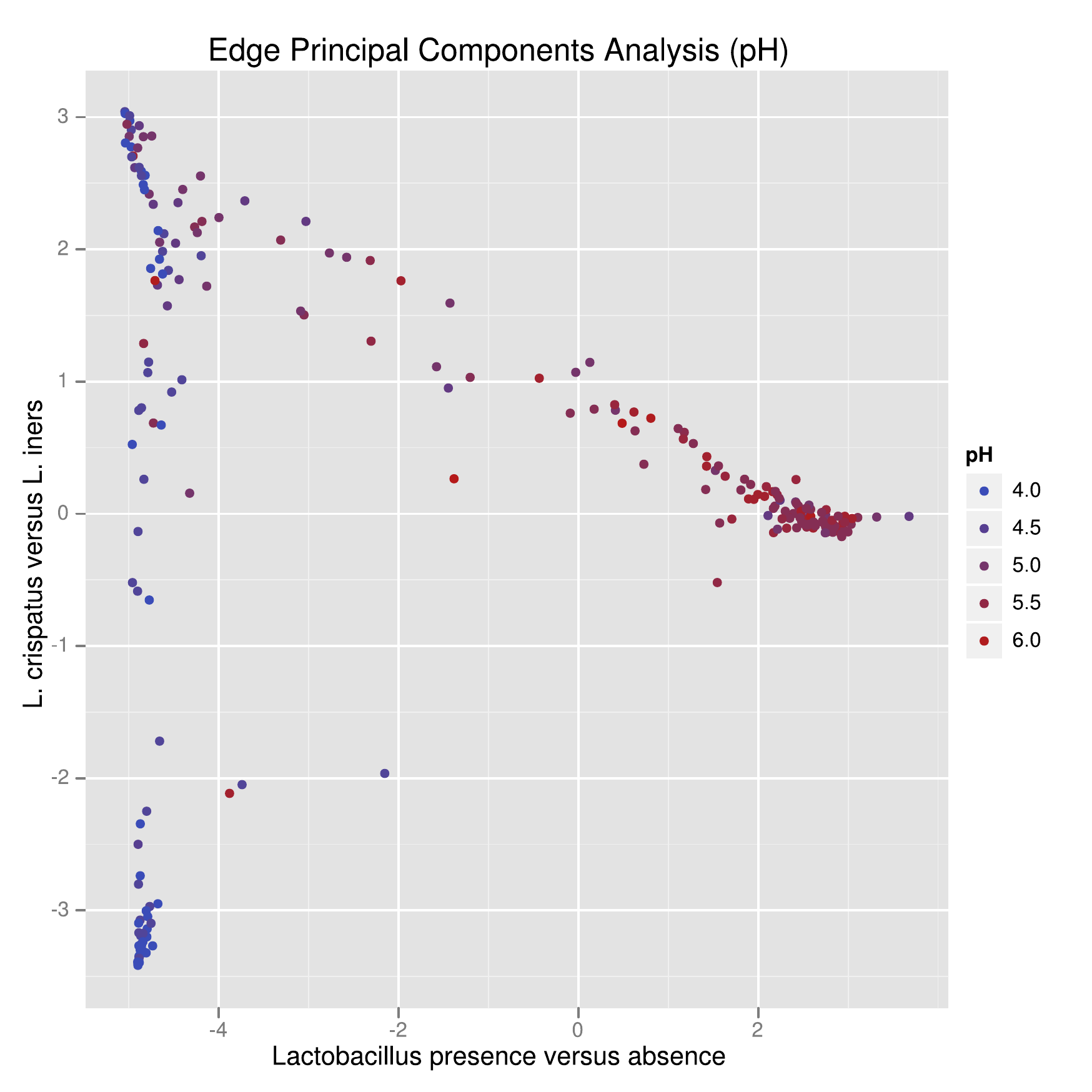}}
\end{center}
\caption{
The samples plotted with respect to the first two principal components and colored according to pH.
}
\label{FIGEdgePCAPlotPh}
\end{figure}
}

\begin{document}

\title[Ordination and clustering for phylogenetic placement data]{Edge principal components and squash clustering: using the special structure of phylogenetic placement data for sample comparison}
\author{Frederick A. Matsen and Steven N. Evans}
\date{\today}

\begin{abstract}
Principal components (PCA) and hierarchical clustering are two of the most
heavily used techniques for analyzing the differences between nucleic acid
sequence samples sampled from a given environment.
However, a classical application of these techniques to distances computed between samples can lack transparency because
there is no ready interpretation of the axes of classical PCA plots, and it is difficult to assign any clear intuitive
meaning to either the internal nodes or the edge lengths
of trees produced by distance-based hierarchical clustering
methods such as UPGMA.
We show that more interesting and interpretable results are produced by two new methods that leverage the special structure of phylogenetic placement data.
{\em Edge principal components analysis} enables the detection of important differences between samples that contain closely related taxa. Each principal component axis is simply a collection of
signed weights on the edges of the phylogenetic tree, and these weights
are easily visualized by a suitable thickening and coloring of the
edges.
{\em Squash clustering} outputs a (rooted) clustering tree
in which each internal node corresponds to an appropriate ``average''
of the original samples at the leaves below the node.  Moreover,
the length of an edge is a suitably defined
distance between the averaged samples associated with
the two incident nodes, rather than the
less interpretable average of distances produced by UPGMA.
We present these methods and illustrate their use with data from the microbiome of the human vagina.
\end{abstract}

\maketitle

\section{Introduction}

Microbial sequence data for a given locus
are naturally endowed with a
somewhat hidden special structure:
the phylogenetic relationships between the organisms
represented by each sequence.  That structure can be inferred by
either building  the phylogenetic tree of the sampled sequences from scratch or by
using {\em phylogenetic placement} techniques to assign to each sampled sequence
a location on a previously constructed reference phylogenetic tree.

In 2005, Lozupone and Knight proposed a way to incorporate this hierarchical structure when computing distances between samples. Their
method, {\em unweighted UniFrac} \cite{LozuponeKnightUniFrac05},
was followed by {\em weighted UniFrac} in 2007 \cite{LozuponeEaWeightedUnifrac07}.
A key feature of both distances is that
differences in community structure due to closely related organisms are weighted less heavily than differences arising from distantly related organisms.
The UniFrac methodology has been widely adopted, and the papers describing the UniFrac variants have hundreds of citations as of the beginning of 2011.

Once distances have been computed between samples using UniFrac, these distances are typically fed into general-purpose ordination and clustering methods, such as classical principal components analysis and UPGMA.
Although it is appropriate to apply such techniques
to distance matrices of this sort, the
classical methods do not use the fact that
the underlying distances were calculated in a specific manner.
Consequently, in an application of principal components analysis, it is difficult
to describe what the axes represent.
Similarly, in hierarchical clustering, it is unclear what is driving a certain agglomeration step; although it can be explained in terms of an arithmetic
operation, a certain amount of interpretability in terms of the
original microbial data is lost.

In this paper, we propose ordination and clustering procedures specifically designed for the comparison of microbial sequence samples that do
take advantage of the underlying phylogenetic structure.
The input for these methods are collections of assignments of
sequencing reads to locations on a pre-existing reference
phylogenetic tree: that is, phylogenetic placements.
Our {\em edge principal components analysis} (edge PCA) algorithm applies
the standard principal components construction to a ``data matrix''
generated from the differences between proportions of phylogenetic placements on either side of each internal edge of the reference phylogenetic tree.
Our {\em squash clustering} algorithm is hierarchical clustering with a novel
definition of distances between clusters that incorporates information
concerning how the data sit on the reference phylogenetic tree.

The primary advantage of these new methods is that of transparency --- namely, that the results of the analyses can be readily visualized and understood.
For example, with edge PCA the principal component axes can be pictured directly in
terms of the reference phylogenetic tree, thereby attaching a clearer interpretation
to the position of a data point along that axis.
Edge PCA is also capable of picking up minor --- but consistent --- differences in collections of placements between samples:
a feature that is important
 in our example application.
The squash hierarchical clustering method has the advantage that each
vertex of the clustering tree is associated with a certain natural distribution of mass on the
phylogenetic tree; the length of an edge in the clustering
tree has a simple interpretation
as the distance between the mass distributions associated with the two
incident vertices.

The ordination and clustering methods presented here are implemented in the \guppy\ binary as part of the \pplacer\ package, available on the pplacer website at \pplacerurl.

\section{Results}

\subsection{General setting for methods}

Phylogenetic placement is a way to analyze the results from high-throughput sequencing applied to DNA extracted in bulk from an environmental sample of microbes.
It is simply the assignment of sequencing reads to a ``reference'' phylogenetic tree constructed from previously-characterized DNA sequences; recent algorithms have focused on doing so according to the phylogenetic maximum-likelihood criterion \cite{BergerStamatakisEPA11,MatsenEaPplacer10}.
By fixing a reference tree rather than attempting to build a phylogenetic tree for the sample from scratch, recent algorithms of this type are able to place tens of thousands of query sequences per hour per processor on a reference tree of one thousand taxa (e.g. species), with performance scaling linearly in the number of reference taxa, the number of query sequences, and the length of the query sequences.

A collection of reads placed on a phylogenetic tree may be thought of as a distribution of a unit amount of mass across the tree.
In the simplest setting, for a collection of $N$ placements on a tree each read is given mass $1/N$; that mass is assigned to the most likely position for that read on the tree.
Another option is to distribute the $1/N$ mass for a given read across the tree in proportion the posterior probability of assignment of that read to various positions \cite{MatsenEaPplacer10}.

This mass distribution may be used to produce distances between collections of phylogenetic placements.
Given two samples for a given locus, each sample is placed individually on the phylogenetic tree, and so each
sample is thought of as a distribution of mass on the tree.
The Kantorovich-Rubinstein or ``earth-mover's'' distance may then be used to quantify the difference between those two samples.
This distance is defined rigorously in \cite{EvansMatsenPhyloKR11}, but the idea is simple to explain.
Imagine that the phylogenetic tree is a road network and that
each sample is represented by the distribution of a unit of mass into piles of dirt along this road network.
The distance between two samples is then
defined to be the minimal amount of ``work'' required to move the dirt
in the first configuration to that in the second configuration
(in this context the amount of work needed to move an infinitesimal mass $\delta$ a distance $x$ is defined to be $\delta \cdot x$).
Thus, similar collections of phylogenetic placements result in similar dirt pile configurations that don't require much mass movement to
transform one into the other, while quite different collections of placements require that significant amounts
of mass must move long distances across the tree.
This distance is classical, having roots in 18th century mathematics, and is a generalization of the {\em weighted UniFrac} distance \cite{EvansMatsenPhyloKR11,LozuponeEaWeightedUnifrac07}.
One can perform PCA and clustering on distance matrices derived from these distances, or, as presented next, develop algorithms that work directly on the underlying data.
These novel algorithms are explained in more detail in \emph{Methods}.

\subsection{Edge principal components analysis}

Suppose that each of $S$ samples is encoded by a
 mass distribution on a reference tree with $E$
internal edges.  We distinguish an arbitrary vertex of the tree as
the {\em root} and map each mass distribution to an $E$-dimensional
vector by recording for each internal
edge the difference between the total mass on the
root side of the edge and the total mass on the non-root side of the edge.
This results in an $S \times E$ ``data matrix''.

Edge principal components analysis (edge PCA) applies
the usual principal components procedure of constructing the $E \times E$
covariance matrix
of this data matrix and then calculating its eigenvalues
and their corresponding eigenvectors.

Each eigenvector can be displayed on the tree, because the coordinates
of the eigenvector correspond to internal edges of the tree.
A large entry in an eigenvector corresponding to one of
the bigger eigenvalues identifies an edge across which there is
substantial heterogeneity among the associated set of mass differences
(see Methods).
In our visualization tool, each eigenvector is represented by a single colored and thickened reference tree: the thickness of an edge is proportional to the magnitude of the corresponding entry of the eigenvector and the color specifies
 the sign of that entry (Fig.~\ref{FIGPCOne} and Fig.~\ref{FIGPCTwo}).
 Moreover, we can project each sample onto an eigenvector
 to visualize how the sample is spread out with respect to that ``axis'' (Fig.~\ref{FIGEdgePCAPlot}).

\subsection{Squash clustering}

Squash clustering is hierarchical clustering with
a novel way of calculating distances between clusters.
Rather than taking averages of distances
as is done in average-linkage clustering (also known as UPGMA),
in squash clustering we take distances between averages of samples.
That is, given a collection of mass distributions on the
reference phylogenetic tree,
each of which correspond to a cluster that has been built at some
stage of the procedure, when the procedure merges two clusters
we simply take a weighted average of the two corresponding mass distributions to get the mass distribution that corresponds to the
new, larger cluster (see Methods).
The ``squash'' terminology describes this averaging procedure:
the original mass distributions for a given cluster are stacked on top of one another and then ``squashed'' down to produce a new object with unit total mass. That is, if we merge two clusters
that correspond to sets of $m$ and $n$ original mass distributions
and are represented
by averaged mass distributions $\mu$ and $\nu$, then the new
cluster is represented by the mass distribution
\[
\frac{m}{m+n} \mu + \frac{n}{m+n} \nu.
\]
Equivalently, if the points in the two clusters were originally
represented by the mass distributions $\mu_1, \ldots, \mu_m$
and $\nu_1, \ldots, \nu_n$, respectively, then the two clusters
are now represented by the mass distributions
\[
\mu = \frac{\mu_1 + \cdots + \mu_m}{m}
\]
and
\[
\nu = \frac{\nu_1 + \cdots + \nu_n}{n},
\]
respectively, and the new cluster obtained by merging them is represented
by the mass distribution
\[
\frac{(\mu_1 + \cdots + \mu_m) + (\nu_1 + \cdots + \nu_n)}{m+n}.
\]

Apart from that difference,
the sequence of steps in the algorithm is identical
to that which occur in the
usual agglomerative hierarchical clustering procedure applied
with the Kantorovich-Rubinstein distances between the initial mass distributions
(i.e. those representing the individual samples) as input.
Each step of agglomerative hierarchical clustering is associated with a pairwise distance matrix and the algorithm
proceeds by merging the pair of clusters
that have the smallest distance between them.
As described above, we take the distance between two clusters $A$ and $B$ to be the earth-mover's distance from the average of the mass distributions in $A$ to the corresponding average for those in $B$.
The series of merges in the clustering algorithm determines the
topology of the rooted {\em clustering tree} that the algorithm produces.
Leaves of the tree correspond to individual samples.
Every internal vertex is associated with a cluster (the collection of samples below that vertex) and thus with a distribution of mass on the phylogenetic tree.
The length of an edge between two arbitrary adjacent vertices on the tree
can be computed by using the earth-movers distance
between the distributions of mass corresponding to those vertices.
This edge length calculation gives the resulting trees
an appearance that differs from that
of UPGMA trees (Fig.~\ref{FIGClusterTrees}) because
the lengths of the paths from the root to the various
leaves are no longer all the same (i.e. the tree is
typically not {\em ultrametric}).

The results of a squash clustering procedure are more transparent than the
equivalent run of a distance-based clustering procedure.
Because of the merging process, each step of squash clustering operates
on the \emph{exactly} the same type of mathematical object: a mass distribution
on a phylgenetic tree.
These mass distributions can be visualized, revealing the similarities that are
driving a particular clustering step (Fig.~\ref{FIGSquash}).

In contrast, for UPGMA or other distance-based hierarchical clustering
techniques, the internal nodes are represented by fundamentally different sorts
of objects than the leaves.
The internal nodes for the classical methods are represented by an
agglomeration of points, and hierarchical clustering variants all have
different ways of using the collection of between-point distances to compute
distances between agglomerations of points.
Consequently, it is not possible to find a manifestation of an internal node
(like the equivalent of one of the mass distributions in Figure~\ref{FIGSquash})
where the distances to that manifestation are the distances used to create the
clustering tree.

These internal node visualizations are automatically generated by the software
implementation of the squash clustering algorithm.
An example application of both edge PCA and squash clustering can be done by running our tutorial at
\texttt{http://fhcrc.github.com/microbiome-demo/}.

\subsection{Example application: the vaginal microbiome}

In this section we apply our clustering and ordination methods to pyrosequencing data from the vaginal microbiome.
For this study, swabs were taken from 242 women from the Public Health, Seattle and King County Sexually Transmitted Diseases Clinic between September 2006 and June 2010 of which 222 samples resulted in enough material to analyze
(Srinivasan, S., Hoffman, N. G., Morgan, M. T., F. A. M., Fiedler, T. L., Hall, R. W., Bumgarner, R., Marrazzo, J. M., and Fredricks, D. N., manuscript in preparation).
DNA was extracted and the 16s gene was amplified in the V3-V4 hypervariable region using broad-range primers and sequenced using a 454 sequencer with FLX chemistry.
Sequences were pre-processed using the R / Bioconductor \cite{rmanual, bioc} package \emph{microbiome}.
A custom maximum likelihood reference tree consisting of sequences from RDP \cite{ColeEaRDP2009} and our local collection was built using \emph{RAxML} 7.2.7 \cite{StamatakisRAxML06} using GTR+4$\Gamma$.
Sequences were placed into this tree using \pplacer\ \cite{MatsenEaPplacer10} with the default parameter choices.

\forarxiv{\FIGPCOne}

The principal components for the vaginal samples independently recover previous knowledge about the contribution of certain microbial species to distinct types of vaginal microbial environment.
In a medical setting, the diagnosis of bacterial vaginosis is often done by looking for rod-shaped bacteria that have a positive Gram stain; these are typically \lacto.
The edge principal component algorithm indicates the importance of this genus: the first principal component for the vaginal data set picks out the presence of \lacto\ versus \sneathia\ and \prevotella\ (Fig.~\ref{FIGPCOne}).
The second principal component reveals that important differences between samples exist at the species level.
Indeed, it highlights the substantial amount of heterogeneity between the amount of two \lacto\ species observed: \liners\ and \lcrisp\ (Fig.~\ref{FIGPCTwo}).
% We might hypothesize that this axis is the ``bacterial vaginosis axis,'' where a projection of the samples onto this axis would have BV-negative samples on the left, and BV-positive samples on the right.

\forarxiv{\FIGPCTwo}

The samples form an interesting pattern when plotted on these axes with their diagnostic score (Fig.~\ref{FIGEdgePCAPlot}).
As described above, samples on the left side have \lacto\ and lack \sneathia\ and \prevotella, while those on the right side have the opposite.
Samples on the bottom have lots of \lcrisp\ and a small amount of \liners, while those on the top have the opposite.
A continuum of samples exists from the lower left to the upper left (mixes of the two prominent \lacto\ species) and from the upper left to the right (from \liners-dominant to \sneathia\ and \prevotella), but there is no continuum from lower left to the right (from \lcrisp-dominant to \sneathia\ and \prevotella).
Reviewing the taxonomically classified data from \cite{SrinivasanEaBV11} confirms this pattern.

\forarxiv{\FIGEdgePCAPlot}

The features detected by the edge PCA algorithm correspond to ones that are pertinent to clinical laboratory diagnosis.
As part of the bacterial vaginosis study, these samples were also processed according to traditional diagnostic criteria.
Vaginal samples were classified in the clinical laboratory according to the Nugent score, which quantifies the presence of various morphotypes (i.e. shapes of bacteria) under a microscope after gram staining.
The Nugent score is considered to be a rigorous standard for the diagnosis of bacterial vaginosis (BV); it ranges from 0 (healthy) to 10 (severe BV).
Swabs were also evaluated for pH.

In general, \lacto\ is associated with a low Nugent score and thus a negative BV diagnosis.
More specifically, \lcrisp\ dominated samples are not found to have a high Nugent score (indicating BV), while \liners\ sometimes are, and those on the right side always are.
Coloring the samples according to pH (Supplementary Fig.~\FIGEdgePCAPlotPhRef) shows a similar pattern.
These plots indicate the possibility of a medically relevant difference between these two \lacto\ species.
We emphasize that the PCA was {\bf not} informed of the Nugent score, pH, or the taxonomic classifications.

\forarxiv{\FIGClassicalPCAPlot}

Applying classical PCA to the pairwise distance matrix does reproduce some of the same features (Fig.~\ref{FIGClassicalPCAPlot}).
However, there is no immediate interpretation  of the
axes output by the classical algorithm.
Furthermore, the important difference between the two \lacto\ species is lost.

\forarxiv{\FIGClusterTrees}
Squash clustering was applied to the collection of vaginal samples in our cohort.
As we have already remarked,
because meaningful internal edge lengths can be assigned to the squash clustering tree, it is not ultrametric, whereas the UPGMA tree is (Fig.~\ref{FIGClusterTrees}).
The two tight clusters at the bottom of (a) and (b) contain the
\lacto-dominated vaginal samples seen on the left side of (Fig.~\ref{FIGEdgePCAPlot}~(a)) and correspond to
\liners\ (upper tight cluster) and \lcrisp\ (lower tight cluster).
A more detailed leaf-labeled comparison between the two trees is available in the supplementary material (Fig.~\FIGClusterMunkresRef).

\subsection{Squash clustering simulation study}

It is difficult to find a collection of microbial communities that have a known hierarchical structure, thus simulation was used to validate the effectiveness of the squash clustering methodology.
The simulation process is described in detail in the \emph{Methods} section, but we highlight several important points here.
The primary ingredients for the simulation are a fixed ``clustering tree'' representing the hierarchical relationship between a set of communities and a ``reference tree'' of species as above.
The simulation generates artificial collections of placements on the reference tree for each leaf of the clustering tree.
The success of the clustering algorithms is judged by comparing the original clustering tree to the result of the clustering method applied to the artificial collections of placements.
This accuracy comparison is done using the rooted Robinson-Foulds (RF) metric (\emph{Methods}).

A number of parameters determine the steps in the simulation process.
Every internal node of the clustering tree is associated with a ``reconstructability'' parameter; this parameter determines the level of similarity between descendants of that internal node.
In this simulation, the reconstructability parameter is set to a single value for all internal nodes of the tree.

\forarxiv{\FIGsquashsim}

Our simulations show that squash clustering and UPGMA applied to KR distances perform similarly across a wide range of simulation parameters (Fig.~\ref{FIGsquashsim}).
Not only do the squash clustering and UPGMA methods have similar levels of accuracy, but their results are also topologically quite similar to one another.
Thus squash clustering, with its more transparently meaningful branch lengths, may prove to be an attractive choice for researchers wishing to find hierarchical structure in their data.

\section{Discussion}

\subsection{Generalization and limitations}
The methods described here, although implemented for comparison of microbial communities, may in fact be used in more general settings.
Edge PCA may be used whenever each sample can be represented by a
collection of mass distributed over a common tree structure.
Squash clustering may be applied in any case where there is a  well-defined notion
of the distance between two samples and a well-defined procedure for
averaging two samples to produce another object of the same type.

There are some limitations to the sort of comparisons that can be performed using these methods simply because the underlying data is a collection of phylogenetic placements on a tree.
For example, if a clade of the reference tree is missing, then differences in diversity within that clade are not be accounted for in the comparison.
Such issues will be present whenever a reference tree is being used, whether using phylogenetic placements directly or mapping reads to the tree using BLAST
as a preliminary step in a UniFrac analysis.
This disadvantage is balanced by the advantage of not having to define organismal taxonomic units (OTU's) by clustering, which can be sensitive to methodological parameters \cite{WhiteEaAlignmentClustering10}.

The methods presented here also depend on the number of phylogenetic placements being correlated with the number of organisms of that type found in the sample.
This is not always true.
Loci such as 16s are often sequenced by first amplifying using a polymerase chain reaction with a broad-spectrum primer; this primer may have different efficiencies for different organisms, or may miss certain organisms altogether.
In addition, genetic material extraction efficiency varies by organism \cite{morganEaInVitroSimulatedMetagenome10}.
Nevertheless, the results on this example data using our methods do appear to correspond with non-genetic methods such as morphological comparison (Fig.~\ref{FIGEdgePCAPlot}) and pH (Supplementary Fig.~\FIGEdgePCAPlotPhRef).

\subsection{Relation to previous work}

The work presented here shares some intent with double principal components (DPCoA) analysis as applied to distributions of phylotypes on a phylogenetic tree \cite{BikEaMicrobiotaStomach06,PurdomAnalyzingDataGraphs08}.
The idea of a DPCoA analysis is to perform a principal components analysis on the phylotype abundance table in a way that down-weights differences between species that are close to one another on the phylogenetic tree.
As such, it shares some similarity to doing multidimensional scaling or principal components on the pairwise distance matrix generated by a KR/UniFrac analysis.
It differs from the methods presented here because it only uses the tree in the form of a pairwise distance matrix; consequently it cannot leverage the edge-by-edge structure of the tree as is done here.

There are also some connections with the statistical comparison features of MEGAN \cite{Mitra09} in that we use the structure of a tree as part of a comparative framework.
Our method and the MEGAN method both highlight regions of the tree for which important differences exist between samples.
However, the details are quite different: the ``directed homogeneity'' test
of \cite{Mitra09} employs statistical criteria to decide if two samples have significant differences either across an edge or between daughters of a given edge.
The edge PCA algorithm, on the other hand, does not attempt to make
hypothesis-testing statistical statement and it uses all of the samples available simultaneously.
In addition, MEGAN performs analysis on taxonomic trees rather than phylogenetic ones.

\subsection{Future work}

The basic step of the edge principal components method--- transforming phylogenetic placement samples into vectors indexed by the edges of the tree--- is general and can be applied in a number of contexts.
In this paper, we followed this transformation with an application of principal components analysis, but many other options are possible.
Next we will apply classical supervised learning techniques to similarly transformed data.

\section{Acknowledgments}

This work would not have been possible without an ongoing collaboration with David Fredricks, Noah Hoffman, Martin Morgan, and Sujatha Srinivasan at the Fred Hutchinson Cancer Research Center.
The authors are grateful for early feedback on this work from Jonathan Eisen and Aaron Darling.
Aaron Gallagher co-developed the \guppy\ framework and helped write the simulation code to validate squash clustering.
Graphics made with ggplot2 \cite{ggplot2Book} and the archaeopteryx tree viewer (http://www.phylosoft.org/archaeopteryx/).
The first author was supported in part by NIH grant HG005966-01, and the second author was supported in part by NSF grant DMS-09-07630.

\section{Author contributions}

F.A.M. conceived of the project and designed and implemented the algorithms.
S.N.E. helped build mathematical foundations and made connections with
the Kantorovich-Rubenstein distance.
F.A.M. and S.N.E. wrote the manuscript.

\notforarxiv{
\clearpage
\section{Figure legends}

\FIGPCOne
\FIGPCTwo
\FIGEdgePCAPlot
\FIGClassicalPCAPlot
\FIGClusterTrees
\FIGsquashsim
\FIGSquash

\clearpage
}

\section{Methods}

A probability measure on the reference phylogenetic tree is obtained from a collection of sequence reads as follows.
A given read can be assigned to the phylogenetic tree in its maximum likelihood or maximum posterior probability location using the phylogenetic likelihood criterion to obtain a ``point placement.''
A point placement can be thought of as a probability measure with all of the mass concentrated at the best attachment location.
Alternatively, one can express uncertainty in the optimal location by spreading the probability mass according to posterior probability (assuming some priors) or ``likelihood weight ratio''; see \cite{MatsenEaPplacer10} for details.
In either case, each read is thought of as a probability measure on the reference phylogenetic tree.
A probability measure for a collection of reads can be obtained by averaging the measures for each read individually (that is, by constructing the probability
measure that is the mixture of the probability measures
for each read in which each such measure is given an equal weight).

\subsection{Edge principal components analysis}

Begin with a phylogenetic tree $T$ and probability measures $P_1,\dots, P_S$ on $T$, each of which comes from an assignment of
the reads in one of $S$ samples to the phylogenetic tree,  as described above.
If $T$ is not already rooted at some vertex, pick an arbitrary vertex to be the root.
Removing a given internal edge $e$ from the tree splits $T$ into two components: $\Tplus(e)$ containing the root and $\Tminus(e)$ without.
For a probability measure $P$ on $T$, define the  corresponding
{\em edge mass difference}
\[
\delta_P(e) = P(\Tplus(e)) - P(\Tminus(e)).
\]
Suppose that $T$ has $E$ internal edges.
The {\em edge mass difference matrix} $\Delta$ is the $S \times E$ matrix that
has the vectors of edge mass differences for the successive samples as its rows.
Edge principal components analysis is then performed by first deriving the $E \times E$ covariance matrix $\Sigma$ from the matrix
$\Delta$ of ``observations'' followed by computing
the $E$ eigenvectors of $\Sigma$ ordered by decreasing
size of eigenvalue.

Each resulting eigenvector is a signed weighting on the internal edges of the tree, and these weightings may be used to highlight those edges of the tree for
which there is substantial between-sample heterogeneity in the masses assigned
to the two components of the tree defined by the edge.
Indeed, recall the variational characterization of the eigenvectors
$v_1, \ldots, v_E$ of an $E \times E$ non-negative definite matrix
$M$ listed in order of decreasing eigenvalue:
\begin{eqnarray*}
  v_1 & = & \argmax_{||v||=1} \langle v , Mv \rangle \\
  v_2 & = & \argmax_{||v||=1, v \perp v_1} \langle v , Mv \rangle \\
  && \\
  && \cdots \\
  v_E & = & \argmax_{||v||=1, v \perp \{v_1, \ldots, v_{E-1}\}} \langle v , Mv \rangle,
\end{eqnarray*}
where $\|v\|$ is the usual Euclidean length of the vector $v$,
$\langle v, w \rangle$ is the usual Euclidean inner product of
the vectors $v$ and $w$, and $v \perp \{v_1, \ldots, v_k\}$ indicates
that $v$ is perpendicular to each of the vectors $v_1, \ldots, v_k$.
Thus, an edge that receives a weight with large magnitude from
an eigenvector corresponding to one of the bigger eigenvalues
of the covariance matrix $\Sigma$
may be viewed as an edge across which there are substantial dissimilarities
between samples in the amount of mass placed in the components on
either side of the edge.

When looking at the weight assigned to a single edge in isolation,
only the magnitude of the weight matters and not the sign, because
if $v$ is an eigenvector for a particular eigenvalue, then
so is $-v$.  However, sign matters when comparing the weights
assigned to two or more edges: if the edge mass differences for
two edges are strongly negatively associated, then the corresponding entry
of the covariance matrix will be very negative, and the corresponding
two entries of the eigenvector for a large eigenvalue will
have different signs.

Changing the chosen root from vertex $x$ to vertex $y$
does not affect the eigenvalues or the
magnitudes of the entries in the corresponding eigenvectors, and it
only changes the signs of the entries for the edges between $x$ and $y$.
This may be seen as follows.
Note first that if an edge $e$ is between $x$ and $y$,
then re-rooting flips the sign of $\delta_P(e)$;
whereas, $\delta_P(e)$ is remains the same if
$e$ is not between $x$ and $y$.
Define $K$ to be the diagonal $E \times E$ matrix such that $K_{e,e} = -1$ for edges $e$ on the path between $x$ and $y$, and $1$ otherwise.
Note that $K = K^{-1}$.
The covariance matrix $\Sigma'$ for the re-rooted tree
and that for the original tree are related
by a similarity transformation: $\Sigma' = K \Sigma K$.
Thus, the eigenvalues for $\Sigma$ are the same as those for $\Sigma'$, and $v_k$ is an eigenvector of $\Sigma$ if and only if
$K v_k$ is an eigenvector of $\Sigma'$.

\subsection{Squash clustering}

Given probability measures $P$ and $Q$ on the
rooted tree $T$, the Zolatarev-like $\kr_p$ generalization of the KR distance is defined for $p \geq 1$ as
\begin{equation}
  \kr_p(P,Q) =
    \left[\int_T \left| P(\tau(y)) - Q(\tau(y)) \right|^p \, \lambda(dy)\right]^{\frac{1}{p}},
\label{EQKR}
\end{equation}
where $\lambda$ is the natural length measure on the tree
and $\tau(y)$ is the subtree on the other side of $y$
from the root \cite{EvansMatsenPhyloKR11}.
The classical KR distance is \eqref{EQKR} with $p=1$; this is the value that corresponds with weighted UniFrac.
It is shown in \cite{EvansMatsenPhyloKR11}
that choosing a different root does not change
the distance.
It is also noted there that if $P$ and $Q$ only assign mass to leaves of the tree and $y$ is in the interior of edge $e$ then
\[
\left| P(\tau(y)) - Q(\tau(y)) \right|
=
\frac{1}{2}
\left(
\left| P(\Tplus(e)) - Q(\Tplus(e)) \right|
+
\left| P(\Tminus(e)) - Q(\Tminus(e)) \right|
\right),
\]
furnishing a connection with edge PCA.

At each stage of the squash clustering algorithm
we have a pairwise distance matrix with
rows and columns indexed by the clusters that
have already been made by the algorithm. Initially,
the clusters are just the individual samples and the
entries in the pairwise distance matrix are computed
using the formula \eqref{EQKR}.

Agglomerative hierarchical clustering in general proceeds by
iterating the following sequence of steps
until there is a single cluster and
a corresponding $1 \times 1$ pairwise distance matrix.

\begin{enumerate}
  \item Find the smallest off-diagonal element in the current
  pairwise distance matrix. Say it is the distance between clusters $i$ and $j$.
  \item Merge the $i$ and $j$ clusters, making a cluster $k$.
  \item Remove the $i$th and $j$th rows and columns from the distance matrix.
  \item Calculate the distance from the cluster $k$ to the other clusters.
  \item Insert the distances from $k$ into the distance matrix.
\end{enumerate}

Classical hierarchical clustering methods calculate the distance in step number 4 as some function of the distance matrix.
In particular, average-linkage clustering or UPGMA calculates the distance between two clusters as the average between pairs of items in the clusters.
Thus, if clusters $i$ and $j$ containing
respective numbers of items $a$ and $b$ are merged
to form a cluster $k$ with $a+b$ items, then the average-linkage distance
between another cluster $\ell$ with $c$ items
and the new cluster $k$ is
(writing $d(\cdot, \cdot)$ for the distance between individual items)
\[
\begin{split}
\mathrm{distance}(k,\ell)
& =
\frac{1}{(a+b)c} \sum_{y \in k, z \in \ell} d(y,z) \\
& =
\frac{a}{a+b} \frac{1}{ac} \sum_{w \in i, z \in \ell} d(w,z)
+
\frac{b}{a+b} \frac{1}{bc} \sum_{x \in j, z \in \ell} d(x,z) \\
& =
\frac{a}{a+b} \mathrm{distance}(i,\ell)
+
\frac{b}{a+b} \mathrm{distance}(j,\ell), \\
\end{split}
\]
and so the entries of the updated UPGMA distance matrix
are just suitably weighted averages of the entries of the
previous distance matrix.

\forarxiv{\FIGSquash}

At each stage of squash clustering, on the other hand,
a cluster is associated with a probability measure on the tree $T$.
When two clusters $i$ and $j$ containing
respective numbers of items $a$ and $b$ and associated
with respective probability measures $P$ and $Q$ are
merged to form a cluster $k$, then the
new cluster $k$ is associated
with the probability measure
$\frac{a}{a+b} P + \frac{b}{a+b} Q$ and the distance
from $k$ to some other cluster $\ell$ associated with the
probability measure $R$ is
\begin{equation}
Z_p \left(\frac{a}{a+b} P + \frac{b}{a+b} Q, R\right),
\label{eq:avgsquash}
\end{equation}
which is analogous to the above formula for the UPGMA averaging procedure.
As remarked above, the ``squash'' interpretation of \eqref{eq:avgsquash} comes from recalling that
the probability measures associated with the two clusters are each simple  averages of
all of the measures for the items in the clusters (Fig..~\ref{FIGSquash}).
That is, if  $S_z$ is the probability measure associated with original item $z$, then
\[
P = \frac{1}{a} \sum_{x \in i} S_x
\]
and
\[
Q = \frac{1}{b} \sum_{y \in j} S_y,
\]
and the probability measure associated
with the new cluster $k$ is
\[
\frac{a}{a+b} P + \frac{b}{a+b} Q
=
\frac{1}{a+b} \sum_{z \in k} S_z,
\]
the (unweighted) average of the probability measures in $z$.

A natural question to ask is whether the distance between
a probability measure $R$
and the weighted average of two probability measures $P$
and $Q$ is equal to the similarly weighted average of
the distance between $R$ and $P$ and the distance
between $R$ and $Q$.
The answer is ``no'':
starting from (\ref{EQKR}) we have from the Minkowski inequality that
for $0 < t < 1$:
\[
\begin{split}
Z_p(t & P + (1-t) Q, R) \\
& = \left[\int_T \left|
      t \left(P(\tau(y)) - R(\tau(y))\right)
      + (1-t) \left(Q(\tau(y)) - R(\tau(y))\right)
    \right|^p \, \lambda(dy)\right]^{\frac{1}{p}} \\
& \leq t \left[\int_T \left| P(\tau(y)) - R(\tau(y))\right|^p \, \lambda(dy)\right]^{\frac{1}{p}}
 +  (1-t) \left[\int_T \left| Q(\tau(y)) - R(\tau(y))\right|^p \, \lambda(dy)\right]^{\frac{1}{p}} \\
& = t Z_p(P, R) + (1-t) Z_p(Q, R).
\end{split}
\]
The early iterations of the UPGMA and squash clustering algorithms
can be quite similar because the pairs of objects being merged are close
together relative to their distance to the other objects.
For example, if $p=1$, then
the above inequality is an equality whenever $P(\tau(y)) - R(\tau(y))$
and $Q(\tau(y)) - R(\tau(y))$ have the same sign for all $y \in T$.

\subsubsection{Squash clustering on ultrametric data}

An appealing feature of UPGMA is that if the
pairwise distances which are used to initialize
the algorithm are the leaf-to-leaf distances for an ultrametric rooted tree $T$,
then UPGMA is guaranteed to return $T$.
In this section we show that squash clustering has a similar property
in a simple special case.
This observation complements
the validation work done using simulation to show that squash
clustering does recover hierarchical structure when it is present.

In order to explain the result for squash clustering, we must first review the simple demonstration of the above result for UPGMA.

Imagine that the ultrametric rooted tree $T$
is oriented on the page with the root at the top and the
leaves at the bottom, and for simplicity assume that it is a bifurcating tree.
By the assumption of ultrametricity, all the leaves will
sit on a horizontal line.
Imagine the internal nodes $z_1, \dots, z_m$ of $T$ are listed in order of
increasing distance from the line
so that $z_1$ is the closest.  For simplicity, suppose further that no
two of these distances are equal, so that we don't have to
adopt an arbitrary convention for breaking ties.
Each internal node corresponds to a set of leaves -- namely, those
that are below it.

We proceed inductively to demonstrate that
the merges done by the algorithm reproduce, in order, the sets of leaves
below the internal nodes $z_1, \dots, z_m$ and that the distances
between clusters assigned by UPGMA agree with the original node-to-node
distances in $T$.
The base case is trivial.
Assume the algorithm satisfies the inductive hypothesis for all $z_j$ with $j <
i$.  The two nodes descending from  $z_i$ in $T$
are each an internal node of the form $z_j$ for some $j<k$ or
a single leaf.  Call the two corresponding sets of leaves below these nodes
$A_i$ and $B_i$. By induction, $A_i$ and $B_i$ are present among
the clusters that have been constructed by UPGMA
after the $(i-1)^{\mathrm{st}}$ merge.
The distance in $T$ between any pair of leaves
$(x,y)$ with $x \in A_i$ and $y \in B_i$
is the same.
By construction, the UPGMA distance between $A_i$ and $B_i$,
\[
(\# A_i)^{-1} (\# B_i)^{-1}
\sum_{x \in A_i, y \in B_i} d(x,y),
\]
is equal to the distance between any two such leaves $x$ and $y$.
Furthermore, the UPGMA distance between $A_i$ (resp. $B_i$) and any
other cluster $C_i$ present after $(i-1)$ UPGMA merges  is equal
to the common distance in $T$ between any leaf in $A_i$ (resp. $B_i$) and any
leaf in $C_i$.  Moreover, by the definition of $z_i$,
this common distance is greater than the
the UPGMA distance between the clusters $A_i$ and $B_i$.
It is now clear that the $i^{\mathrm{th}}$ merge of UPGMA
merges the clusters $A_i$ and $B_i$ to produce a cluster
that coincides with the set of leaves below $z_i$ in $T$
and that the updating of distances maintains the agreement
between node-to-node distances in $T$ and UPGMA cluster-to-cluster distances.

A similar argument leads to an analogous statement for squash clustering.
Again, assume that the reference tree $T$
is an ultrametric rooted tree.
For each leaf $\ell$, assume that there is a single sample $S_\ell$
consisting of a single read mapped to $\ell$.
We will show that in this case both squash clustering and UPGMA applied to KR
$Z_1$ distances return the reference tree $T$ as the clustering tree.

First note that the $Z_1$ distance between the two samples $S_\ell$ and $S_{\ell'}$ is
simply the distance on the tree between the leaves $\ell$ and $\ell'$.
These distances are ultrametric by assumption.
Thus, UPGMA run with KR
distances will return $T$ as the clustering tree in this case.

Further, squash clustering and UPGMA start with the same clusters (each read in
a cluster by itself), every cluster is trivially the set of leaves
below a node of the reference tree
$T$, and the distances between clusters are the same for the two methods.
Suppose, then, that after some number of iterations
of the two methods we are still in a situation
where the two methods have the same clusters available to merge,
these clusters are disjoint sets of leaves below nodes of $T$,
and the distances between the clusters available to merge
are the same for the two methods.

Call the available clusters $C_1, \ldots, C_m$.
By definition, squash clustering and UPGMA
will merge the same pair of clusters --
say, without loss of generality, $C_1$ and $C_2$.
The $Z_1$ squash clustering distance is the optimal transport
(earth movers') distance
between the probability measure that puts mass $(\# C_1 + \# C_2)^{-1}$
at each leaf of $C_1 \cup C_2$
and the probability measure that puts mass $(\# C_i)^{-1}$
at each leaf of $C_i$ for $i > 2$.
Because, as we remarked above,
$d(x',y') = d(x'',y'')$ for any
$x',x'' \in C_1 \cup C_2$ and $y',y'' \in C_i$, $i > 2$,
the optimal transport distance is necessarily this common value.
Thus, the updating of the distances between the clusters
available for merging is the same for the two methods.
Therefore, by induction, the trees produced by
the two methods will be the same and will coincide with
the tree $T$.

\subsection{Simulation methodology for clustering validation}

In this section we present methodology for making artificial ``samples'' that are hierarchically related.
These are then used to compare squash clustering to UPGMA.
The code for these simulations can be found on the \textsf{commiesim} branch of \pplacer\ at \\
\texttt{http://github.com/matsen/pplacer/tree/commiesim}.

Start with a true ``clustering tree'' $C$: the tree of communities on which we are simulating.
Let $T$ be a phylogenetic ``reference'' tree of the organisms of interest: the phylogenetic tree of the actual species from which the simulated placements will be drawn.
Write $L$ for the set of leaves of $T$.
Before describing the simulation
we recall some standard terminology.
A {\em split} of $T$
is the partition of the leaves $L$ induced by an edge of $T$:
it consists of the two subsets of $A,B$ of $L$ that
are on either side of the edge.  We have $A \cup B = L$ and
$A \cap B = \emptyset$, and we use the notation $A|B$ to
denote that the subsets $A$ and $B$ form a split.

The first step of simulation assigns subsets of $L$
to the leaves of the clustering tree $C$.
The elements of each such subset
are the organisms found in that particular ``community'';
community will then be used to generate simulated placements by sampling some number of members of the community with replacement.
For example, suppose that a leaf $x$ of the clustering tree $C$ is associated with the set $S$ of leaves of the reference tree $T$;
to generate a sampled collection of placements
for $x$ we first sample from $S$ with replacement.
The resulting multi-set of leaves of the reference tree $T$ is made into a collection of placements by turning each element into a placement
consisting of a unit point mass at the given leaf of the reference tree.

These simulated collections of placements are then used
to reconstruct the clustering tree by applying
both squash clustering or UPGMA on the KR distances.

Subsets of the leaf set $L$ of $T$
are assigned to leaves of the clustering tree $C$
by a recursive procedure that proceeds down the clustering tree
beginning with the root $r$.
At each stage there is a
current internal node $t$ of $C$ and a
set of leaf sets $J_t$ associated with $t$.
The recursion is initialized with $J_r = \{ L \}$.
We proceed down the tree $C$ from a node $t$ in two stages:
we first split the set of subsets $J_t$
and then assign some of these subsets to each child of $t$.

The splitting stage is done by selecting splits
of $T$ and using them to cut apart the leaf subsets.
For example, suppose that $J_t = \{S_1,\ldots , S_k\}$ is the set
of subsets of $L$ associated with
the internal node $t$ of $C$ that we are currently processing.
We select an ``effective'' split $A|B$ of $T$, i.e. one such that $A \cap S_i$ and $B \cap S_i$ are non-empty for some $i$.
Applying this split produces the new collection
of leaf subsets $\{S_1, \ldots, A \cap S_i, B \cap S_i, \ldots, S_k\}$.
Each one of the $S_j$ corresponds to a connected region of the reference tree $T$, and applying an effective split corresponds to disconnecting
one of those regions by cutting an edge of $T$.
In the simulation, we sample an integer $e$ from a Poisson distribution with mean $\mu$ and
then sample $e$ effective splits uniformly with replacement
from the set of all effective splits for the subsets in $J_t$.
We apply those splits successively as above to split the subsets in $J_t$.
This splitting produces a new set of leaf subsets
that we call $K_t$.

Next, for each child of the current internal node $t$,
we select a subset of $K_t$ of size $n$ to pass on to the child.
We do this in such a way that $q$ of the subsets selected
are the same for each child, while the remaining $n-q$
are selected independently of the corresponding selections
for the other children.  Here $n$ is a fixed parameter
and $q$ is a realization of a binomial distribution
with number of trials $n$ and success parameter $0 \le r_t \le 1$.
The ``reconstructability parameter'' $r_t$ determines the level of similarity between the children of $t$:
for internal nodes with high $r_t$ the subsets assigned to its
children will be quite similar, while for those with low $r_t$
the subsets will tend to be different.

More specifically, suppose that
the children of $t$ are the nodes $t_1, \cdots, t_\ell$.
We first sample $q$ elements from $K_t$ with replacement to make a set $M$ of subsets of $L$ with at most $q$ elements.
% {\tt Am I right here, or is $M$ really a multi-set? Is the same true for the $S_j$?}
% Yes, indeed these are sets, not multi-sets.
Next, for $1 \le i \le \ell$, we sample $n-q$ elements from $K_t$ with replacement to make a set $L_i$ of subsets of $L$
with at most $n-q$ elements.
Then, $J_{t_i}$, the set of subsets associated
with the node $t_i$, is defined to be the set $M \cup L_i$.
By recursing in this fashion, every node $t$ of the clustering tree $C$ is assigned some set $J_t$ of subsets of the set of leaves
$L$ of the reference tree $T$.
For each leaf $t$ of the clustering tree, placements are simulated as described above from the set of leaf subsets $J_t$.

For the study reported in Figure~\ref{FIGsquashsim}, the following parameters were used.
The clustering tree $C$ was, in the usual bracketing
notation for binary rooted trees, the tree $((a,(b,c)),(d,(e,f)))$.
The reference tree $T$ was the tree for microbes in the vaginal environment used in the rest of the paper.
500 trials were performed for every parameter setting, and 100 placements were generated for each clustering leaf of each trial.
The mean number of cuts $\mu$ was set to 10, and the number of sets selected $n$ was set to 5.
The reconstructability parameters $r_t$ for all internal nodes were set to the value specified in the panel label of the figure.

The Robinson-Foulds (RF) metric \cite{robinson1981comparison} of two trees $T$ and $S$ was computed as half the size of the symmetric difference of the split-set of $T$ and that of $S$.
Because the classical RF distance is calculated on unrooted trees,
while the clustering trees in the study are rooted,
we attached a fictitious ``root leaf'' to the root before calculating RF distances
to account for the position of the root.
We call the resulting quantity the \emph{rooted Robinson-Foulds distance}.
For a bifurcating tree on six leaves such as $C$, the maximal rooted RF distance is four.

\bibliographystyle{plain}
\bibliography{edge_squash}

\newpage

% Supplementary figure labeling, i.e. S1, S2, \dots
\setcounter{figure}{0}
\setcounter{table}{0}
\makeatletter
\renewcommand{\thefigure}{S\@arabic\c@figure}
\renewcommand{\thetable}{S\@arabic\c@table}
\makeatother

\section{Supplementary Figures}

\forarxiv{
\FIGClusterMunkres
\FIGEdgePCAPlotPh
}

\end{document}